\documentclass[reprint,floatfix,twocolumn,10pt]{revtex4-1} 

\usepackage{graphicx}
\usepackage{amsmath}
\usepackage{amsfonts}
\usepackage{amssymb}
\usepackage{amsthm}
\usepackage{hyperref}
\usepackage{color}

\newcommand{\bra}[1]{\ensuremath{\left\langle#1\right|}}
\newcommand{\ket}[1]{\ensuremath{\left|#1\right\rangle}}

\newcommand{\ketbra}[2]{\ket{#1}\!\!\bra{#2}}

\newtheorem{theorem}{Theorem}
\newtheorem{lemma}{Lemma}
\newtheorem{corollary}{Corollary}

\graphicspath{ {./Figures/} }

\newcommand{\eqnref}[1]{Eq.~(\ref{#1})}

\newcommand{\onehalf}{\frac{1}{2}}
\newcommand{\onethird}{\frac{1}{3}}
\newcommand{\oned}{\frac{1}{d}}

\newcommand{\ztwo}{\mathbb{Z}_2}
\newcommand{\ztwotwo}{(\mathbb{Z}_2)^2}
\newcommand{\ztwothree}{(\mathbb{Z}_2)^3}
\newcommand{\ztwom}{(\mathbb{Z}_2)^m}

\newcommand{\inv}{^{-1}}
\newcommand{\bitens}{\hat{\tau}_2}
\newcommand{\tritens}{\hat{\tau}_3}
\newcommand{\tristate}{\ket{\psi(\tau_3)}}
\newcommand{\Gthird}{G_{\onethird}}

\begin{document}

\title{Latent Computational Complexity of Symmetry-Protected Topological Order \\ with Fractional Symmetry}

\author{Jacob Miller}
\email{jmjacobmiller@gmail.com}
\author{Akimasa Miyake}
\email{amiyake@unm.edu}
\affiliation{Center for Quantum Information and Control, Department of Physics and Astronomy, University of New Mexico, Albuquerque, NM 87131, USA}

\begin{abstract}

An emerging insight is that ground states of symmetry-protected topological orders (SPTO's) possess latent computational complexity in terms of their many-body entanglement. By introducing a fractional symmetry of SPTO, which requires the invariance under 3-colorable symmetries of a lattice, we prove that every renormalization fixed-point state of 2D $\ztwom$ SPTO with fractional symmetry can be utilized for universal quantum computation using only Pauli measurements, as long as it belongs to a nontrivial 2D SPTO phase. Our infinite family of fixed-point states may serve as a base model to demonstrate the idea of a ``quantum computational phase'' of matter, whose states share universal computational complexity ubiquitously.

\end{abstract}

\maketitle

\section{Introduction}
\label{sec:introduction}

Understanding the varied correspondence between quantum entanglement and quantum computation is one of the leading goals of quantum information science. Measurement-based quantum computation (MQC) \cite{raussendorf2001one, raussendorf2003measurement, jozsa2005introduction}, where computation is driven by single-spin measurements on a many-body resource state, lets us study this correspondence directly, in terms of the computations achievable with a fixed resource state. Of particular interest are the universal resource states, whose many-body entanglement lets us implement any quantum computation efficiently \cite{nest2006universal, nest2007fundamentals, chen2010quantum}. In trying to characterize the entanglement found in universal resource states, researchers have developed a long list of examples, from the 2D cluster state \cite{briegel2001persistent, zhou2003quantum} and certain tensor network states \cite{verstraete2004valence, nest2006universal, nest2007fundamentals, gross2007novel, gross2007measurement, chen2009gapped, cai2010universal}, to condensed matter models such as 2D Affleck-Kennedy-Lieb-Tasaki (AKLT) states \cite{miyake2011quantum, wei2011affleck, darmawan2012measurement, wei2013quantum, wei2015universal} and renormalization fixed-point states of interacting bosonic quantum matter \cite{miller2016hierarchy, nautrup2015symmetry}.

An emergent insight from these examples has been the utility of symmetry-protected topological order (SPTO), a form of quantum order arising from nontrivial many-body entanglement protected by a symmetry \cite{pollmann2010entanglement, gu2009tensor, kitaev2009periodic, ryu2010topological, chen2011classification, schuch2011classifying, pollmann2012symmetry, chen2012symmetry, chen2013symmetry}. This insight has led researchers to investigate a general correspondence between SPTO and MQC, with the ultimate aim of discovering a ``universal computational phase" of quantum matter. In such a phase, the constituent states' SPTO and symmetry alone structure them as universal resource states. While this approach has uncovered increasingly general single-qubit computational phases in 1D spin chains \cite{brennen2008measurement, doherty2009identifying, skrovseth2009phase, bartlett2010quantum, miyake2010quantum, else2012symmetryprl, else2012symmetrynjp, miller2015resource, prakash2015ground, raussendorf2017symmetry, stephen2017determining}, much less is known in the computationally important setting of 2D spin systems outside of variously perturbed phases containing the cluster state \cite{son2011quantum, kalis2012fate, orus2013bounds, fujii2013measurement, wei2014transitions}. This disparity comes both from the increased complexity present in 2D many-body systems, as well as the existence of physically distinct forms of 2D SPTO with different operational capabilities \cite{miller2016hierarchy}. For these reasons, we have yet to figure out even a base model for realizing the idea of a universal computational phase within the framework of SPTO.

Here, our key starting point is to focus on 2D model states representing renormalization group (RG) fixed-point states of SPTO. As described in detail in Appendix~\ref{sec:renormalization}, these ``3-cocycle states'' \cite{chen2013symmetry} define a coarse-grained, yet infinitely large, family of representative wavefunctions which are macroscopically distinct regarding their SPTO. In addition to the standard abelian, on-site symmetry groups $G=\ztwom$, we introduce an additional fractional $\tfrac{1}{3}$ symmetry of 2D lattice geometry, where symmetry operators are applied to only a certain fraction of spins on a 3-colorable lattice. It turns out that this fractional symmetry is powerful enough to establish a \textit{one-to-one} correspondence between the computational universality of these states for MQC and the non-triviality of SPTO phases they represent in terms of cohomology classes. Our findings form compelling evidence pointing towards universal computational 2D phases among general fractionally symmetric SPTO states, with the expectation that an operational analysis of RG flows may be feasible along the same lines as the aforementioned success of RG methods in 1D spin chains.

\section{Background}
\subsection{Measurement-based Quantum Computation}

Measurement-based quantum computation (MQC) utilizes an entangled many-body resource state to perform quantum computation via local measurements on single lattice sites. An MQC protocol is adaptive if the choice of measurement basis depends on previous measurement outcomes. A universal resource state is one which allows any unitary quantum circuit to be efficiently implemented using single-site measurements. 

\begin{figure}[t]
  \centering
  \includegraphics[width=0.5\textwidth]{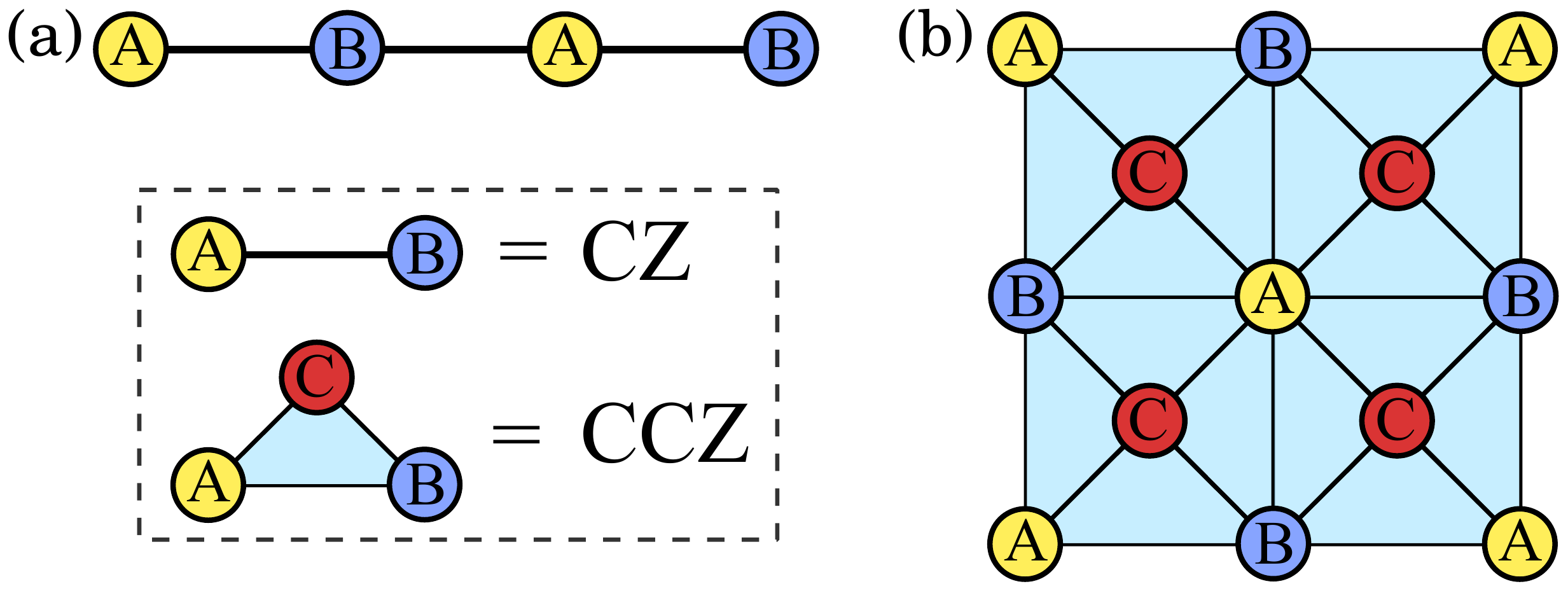}
  \caption{The 1D cluster state $\ket{\psi_{1C}}$ (a), and 2D Union Jack state $\ket{\psi_{UJ}}$ (b), canonical examples of the entangled many-body states we investigate. (a) The 1D cluster state is formed from qubit $\ket{+}$ states (with $\ket{+}:=\frac{1}{\sqrt{2}}(\ket{0}+\ket{1})$) on a 1D spin chain, which are entangled with nearest-neighbors $CZ$ gates acting as $CZ \ket{\alpha,\beta} = (-1)^{\alpha\cdot\beta} \ket{\alpha,\beta}$. (b) The Union Jack state is obtained from $\ket{+}$ states on a 2D Union Jack lattice, which are entangling with nearest-neighbor triple $CCZ$ gates acting as $CCZ \ket{\alpha,\beta,\gamma} = (-1)^{\alpha\cdot\beta\cdot\gamma} \ket{\alpha,\beta,\gamma}$. Both $\ket{\psi_{1C}}$ and $\ket{\psi_{UJ}}$ possess distinctive ``fractional symmetries", leaving them invariant under $X$ applied to all qubits on sites of a single color ($A$, $B$, or $C$). Replacing the $(d-1)$-controlled $Z$ gates by unitaries $U(\omega_d)$ parameterized by $d$-cocycles of a group $G$, we obtain the cocycle states of \cite{chen2013symmetry}. Cocycle states with fractional symmetry can be graphically represented by expanding every vertex into a collection of virtual qubits, and expanding every entangling gate into a product of $CZ$'s or $CCZ$'s (see Figure~\ref{fig:classification}).}
  \label{fig:1C_UJ}
\end{figure}

While MQC has historically focused on the 2D cluster state \cite{briegel2001persistent}, which has a peculiar nature regarding SPTO (see Appendix~\ref{sec:cohomology}), we are more interested here in its 1D spin chain cousin and the Union Jack state of \cite{miller2016hierarchy} (see Figure~\ref{fig:1C_UJ}). Within MQC, the 1D cluster state can implement all single-qubit operations, while the Union Jack state is universal using only Pauli measurements, a property called Pauli universality.

\subsection{Symmetry-Protected Topological Order}

Symmetry-protected topological order (SPTO) is a quantum phenomenon in many-body systems with global symmetry $G$, which will always be abelian here. An SPTO phase is the collection of all many-body states connected to some fiducial short-range entangled state using only constant depth quantum circuits built from constant range, symmetry-respecting gates. The trivial SPTO phase is the unique phase containing unentangled product states. Nontrivial SPTO consequently represents a form of persistent many-body entanglement, protected by a symmetry group $G$.

SPTO phases can be classified using group cohomology theory \cite{chen2013symmetry}. For 2D states, SPTO phases relative to $G$ correspond to elements of the third cohomology group of $G$, $\mathcal{H}^{3}(G,U(1))$. We can analyze $\mathcal{H}^{3}(G,U(1))$ using 3-cocycles, complex-valued functions $\omega_3(g_1,g_2,g_3): G^3\to U(1)$ which satisfy the condition $\partial_3\omega_3(g_0,g_1,g_2,g_3) := \omega_3(g_1,g_2,g_3)\,\allowbreak \omega_3^*(g_0g_1,g_2,g_3)\,\allowbreak \omega_3(g_0,g_1g_2,g_3)\,\allowbreak \omega_3^*(g_0,g_1,g_2g_3)\,\allowbreak \omega_3(g_0,g_1,g_2) = 1$, for all $g_0,g_1,g_2,g_3\in G$. Each 3-cocycle $\omega_3$ lies in a unique ``cohomology class", $[\omega_3]_G\in \mathcal{H}^{3}(G,U(1))$, where the cohomology class of the function $\omega_3(g_1,g_2,g_3) = 1$ is the trivial SPTO phase. In general $d\geq1$ spatial dimensions, SPTO phases are classified by $\mathcal{H}^{d+1}(G,U(1))$.

\subsection{Cocycle States}

While the correspondence between SPTO phases and cohomology classes may appear obscure, it lets us construct useful SPTO fixed-point states using the cocycle state model of \cite{chen2013symmetry}. This model converts abstract $d$-cocycles $\omega_d$ of $G$ into $d$-body unitary gates $U(\omega_d)$, which then form many-body states $\ket{\psi(\omega_d)}$ in $d-1$ spatial dimensions. These states have global symmetry $G$, and belong to the SPTO phase associated with $[\omega_d]_G$. We discuss only the 2D case ($d=3$), but this method extends to any $d\geq1$ spatial dimensions.

For any $G$, $\ket{\psi(\omega_3)}$ is made of $|G|$-dimensional qudits on a 2D lattice $\Lambda$ without boundaries. On-site symmetry operators $X_g$ act in a generalized computational basis as $X_g\!\ket{h} = \ket{gh}$, $\forall g,h\in G$. When $G=\ztwom$, a generating set for $G$ (explained below) lets us represent each qudit as $m$ ``virtual" qubits, on which $X_g=\bigotimes_{i=1}^m (X_i)^{g_i}$. We visualize these qubits stacked in vertical layers, from $i=1$ (top) to $i=m$ (bottom). The state $\ket{+_G}=\ket{+}^{\otimes m}$ is the unique $+1$ eigenstate state of every $X_g$.

$\omega_3$ sets the eigenvalues of our entangling unitary $U(\omega_3)$, as $U(\omega_3) = \sum_{g,h,f\in G} \omega_3(g,g\inv h,h\inv f) \ketbra{g,h,f}{g,h,f}$. We form $\ket{\psi(\omega_3)}$ from $\ket{+_G}$ states on every vertex of a 3-colorable lattice, with $U(\omega_3)$ (or $U(\omega_3)^\dagger$) applied to all nearest-neighbor triples of qudits $\Delta_3$. The three arguments $g,h,f$ match the three qudits in $\Delta_3$ according to their lattice colors. Overall, $\ket{\psi(\omega_3)} = \big(\prod_{\Delta_3\in\Lambda} U(\omega_3)_{\Delta_3}^{\pm1}\big) \ket{+_G}^{\otimes n}$, where the alternation of $U(\omega_3)$ and $U(\omega_3)^\dagger$ is described in \cite{chen2013symmetry}.

The 1D cluster state and Union Jack state are both $G=\ztwo$ cocycle states, with respective cocycles $\omega_2^{(1C)}(g,h) = (-1)^{g\cdot h}$ and $\omega_3^{(UJ)}(g,h,f) = (-1)^{g\cdot h\cdot f}$ (c.f. Appendix~\ref{sec:cohomology} of \cite{miller2016hierarchy}). However, these states both possess additional ``$\oned$'' fractional symmetry, arising from $X_g$ applied to spins of a single vertex color on a $d$-colorable lattice. As we show below, this fractional symmetry is connected to each cocycle being a $d$-linear function, something we define explicitly for $d=3$.

A function $\tau_3(g,h,f):G^3\to U(1)$ is 3-linear (trilinear) when it satisfies $\tau_3(gg',h,f) = \tau_3(g,h,f)\tau_3(g',h,f)$, and similarly for its other two arguments. Every trilinear function is a 3-cocycle, but one possessing additional algebraic structure. This lets us efficiently describe $\tau_3$ by choosing a generating set for $G=\ztwom$, namely a collection of $m$ elements $\{e_i\}_{i=1}^m\subseteq G$ by which every $g\in G$ is $g=\prod_{i=1}^m (e_i)^{g_i}$ for a unique choice of binary coordinates $g_i$. Given a fixed generating set, we have $\tau_3(g,h,f)=(-1)^{\sum_{i,j,k=1}^m \tritens(i,j,k)\cdot g_i\cdot h_j\cdot f_k}$, where $i,j,k$ index the generators of $\ztwom$, and $\tritens(i,j,k)$ is a binary ``component" of $\tau_3$ encoding the value of $\tau_3(e_i,e_j,e_k)$. These components form an $m\times m\times m$ binary tensor $\tritens$, whose transformation under index-dependent changes of generating set will concern us below. We can similarly define 2-linear (bilinear) functions $\tau_2(g,h)$, described by $m\times m$ binary component matrices $\bitens(i,j)$. For more information on group cohomology, the cocycle state model, and the formulation of so-called stabilizer states as examples of cocycle states, see Appendix~\ref{sec:cohomology}.

\section{Cocycle States with Fractional Symmetry}

Given the fractional symmetry of the 1D cluster state and Union Jack state, we ask how this symmetry orders the entanglement of general many-body states. Our main results form a largely exhaustive answer to this question for 1D 2-cocycle states and 2D 3-cocycle states. We first show that any $\oned$-symmetric cocycle state with $d=2$ or 3 and $G=\ztwom$ is either a trivial product state, or is reducible by local operations to several disjoint copies of the 1D cluster state or the Union Jack state, respectively. For $d=2$, this characterization is complete, in that every nontrivial $\onehalf$-symmetric cocycle state $\ket{\psi(\omega_2)}$ is isomorphic to $r$ copies of $\ket{\psi_{1C}}$, for an $\omega_2$-dependent $r\geq1$. When $d=3$ however, we show that general $\onethird$-symmetric cocycle states with $G=\ztwom$ are isomorphic to $r$ ``irreducible" 3-cocycle states, of which the Union Jack state is the simplest. This proves that all nontrivial 3-cocycle states with $\onethird$-symmetry and $G=\ztwom$ are Pauli universal MQC resource states, identifying a robust correspondence between fractional symmetry and the utility of many-body states for quantum computation.

We first characterize the algebraic properties of cocycle states with $\oned$-symmetry. We show that for $d=2$ and 3, $d$-cocycle states with $\oned$-symmetry are precisely those generated by $d$-linear functions (Lemma~\ref{lem:one}).

\begin{lemma}
\label{lem:one}
    Let $\ket{\psi(\omega_d)}$ be a $d$-cocycle state defined on a $d$-colorable $(d-1)$-dimensional lattice without boundaries, generated by a $d$-cocycle $\omega_d$ with $d=2,3$. $\ket{\psi(\omega_d)}$ is $\oned$-symmetric, i.e. is invariant under the application of $G$ to all sites of any one of the $d$ lattice colors, if and only if it is generated by a unique $d$-linear function $\tau_d$, so that $\ket{\psi(\omega_d)} = \ket{\psi(\tau_d)}$.
\end{lemma}

\noindent Lemma~\ref{lem:one}'s statement can be generalized to arbitrary $d$, but due to our focus on low-dimensional MQC resource states, this generalized version remains a conjecture. Proving that $d$-linear cocycle states possess $\oned$-symmetry is trivial, so we focus on the reverse implication. Our proof analyzes the action of fractional symmetry operators on local regions of a $d$-cocycle state $\ket{\psi(\omega_d)}$, and iteratively builds up necessary conditions for $\ket{\psi(\omega_d)}$ to possess $\oned$-symmetry. This shows that $\omega_d$ is the product of a unique $d$-linear $\tau_d$ with additional terms acting on the boundaries of our system, proving our result. The full proof of Lemma~\ref{lem:one} is given in Appendix~\ref{sec:lemone}.

The specification of $d$-linear $\tau_d$'s using component tensors $\hat{\tau}_d$ lets us decompose $U(\tau_d)$ into a product of $d$-qubit component unitary gates, one for each nonzero component of $\hat{\tau}_d$. When $G=\ztwom$ and $d=2$ or 3, these component gates are $CZ$ or $CCZ$, which shows each $\ket{\psi(\tau_d)}$ to be a so-called hypergraph state \cite{rossi2013quantum, guhne2014entanglement, steinhoff2017qudit}. This decomposition of $U(\tau_d)$ into $CZ$ or $CCZ$ gates requires a choice of generating set for each vertex color of our $d$-colorable lattice, with changes of generating set acting as gauge freedoms in the description of $\ket{\psi(\tau_d)}$. We can fix these spurious degrees of freedom by enumerating the local unitary orbits of $\ket{\psi(\tau_d)}$ under color-dependent changes of basis, which reduces to finding a normal form for our component tensor $\hat{\tau}_d$.

For 1D and 2D states, this classification reduces to that of irreducible $\oned$-symmetric cocycle states $\ket{\psi(\gamma_i)}$ (defined below), as given in Theorem~\ref{thm:one}.

\begin{theorem}
\label{thm:one}
Let $\ket{\psi(\tau_d)}$ be a nontrivial $\oned$-symmetric $d$-cocycle state without boundaries in $d-1$ spatial dimensions, with global symmetry $G=\ztwom$ and $d=2,3$. By an appropriate color-dependent change of basis, there is a unique $r$ with $1\leq r \leq m$ such that the nontrivial portion of $\ket{\psi(\tau_d)}$ is isomorphic to $r$ disjoint irreducible $\oned$-symmetric cocycle states, i.e. $\bigotimes_{i=1}^{r} \ket{\psi(\gamma_i)}$.
\end{theorem}

\begin{figure*}[t]
  \centering
  \includegraphics[width=1.00\textwidth]{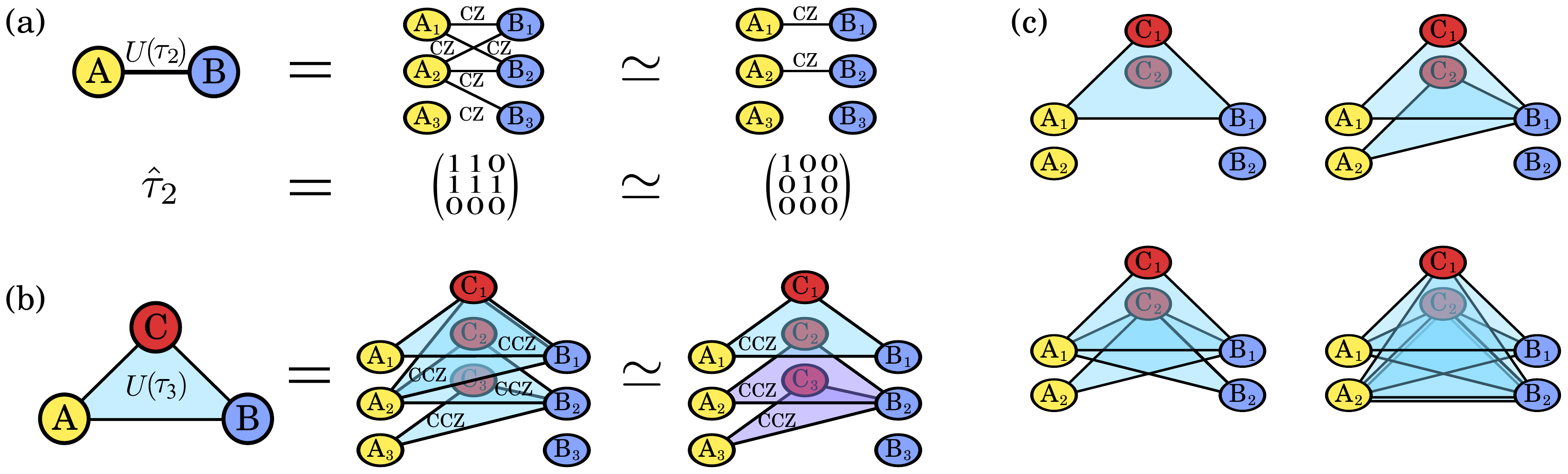}
  \caption{(a) Fixing a $G=\ztwom$ generating set at sites $A$ and $B$ lets us represent the entangling gates $U(\tau_2)$ forming our $\onehalf$-symmetric 2-cocycle state using an $m\times m$ binary component matrix, $\bitens$. Nonzero entries of $\bitens$ give $CZ$ gates between adjacent virtual qubits. Color-dependent changes of generating set (corresponding to color-dependent changes of basis on the single-spin Hilbert spaces), enact Gaussian elimination, reducing $\bitens$ to a diagonal normal form in which our state is simply $r=\textrm{rank}(\bitens)$ disjoint 1D cluster states. Here, $m=3$ and $r=2$. (b) In 2D, we again use color-dependent changes of generating set to simplify our state, but now represent 3-body entangling gates $U(\tau_3)$ as 3-index binary component tensors, $\tritens$ (not shown). Nonzero entries of $\tritens$ give $CCZ$ gates between triples of virtual qubits. Our normal form reduces this state to $r$ disjoint irreducible 3-cocycle states, where again $m=3$ and $r=2$. (c) Representatives of the $\zeta_3(2)=4$ irreducible 3-cocycle states which exist in $G=\ztwotwo$. Theorem~\ref{thm:one} proves that any $\onethird$-symmetric 3-cocycle state with $G=\ztwom$ is either trivial, isomorphic to one of these states (up to permutation of lattice colors), or isomorphic to two disjoint copies of the Union Jack state (the only irreducible state in $\ztwo$). An exhaustive numerical search shows that of the $2^{m^3}=2^{27}$ possible $\onethird$-symmetric cocycle states in $G=\ztwothree$, there exist only $\zeta_3(3)=50$ distinct irreducible states up to local changes of basis. However, a precise classification of irreducible cocycle states is unnecessary for our purposes, since every irreducible state is a Pauli universal resource state (Corollary~\ref{cor:one}).}
\label{fig:classification}
\end{figure*}

We let $\zeta_d(m)$ denote the number of distinct irreducible $d$-cocycle states in $G=\ztwom$, which is calculated using the component tensors $\hat{\tau}_d$. When $d=2$, we reduce $\bitens$ to normal form using color-dependent changes of generating set on lattice colors $A,B$, transforming $\bitens$ to $\chi_A^T \bitens \chi_B$ with invertible binary matrices $\chi_A,\chi_B$. Choosing $\chi_A$ and $\chi_B$ to implement elementary row and column operations, we can transform $\bitens$ into a diagonal normal form using Gaussian elimination. This gives $U(\tau_2)$ as a product of disjoint $CZ$ gates forming $r$ disjoint copies of $\ket{\psi_{1C}}$, with $r$ the rank of $\bitens$ (see Figure~\ref{fig:classification}a). This proves Theorem~\ref{thm:one} for $d=2$, and shows also that $\zeta_2(m)=1$ for all $m$, meaning the 1D cluster state is the unique irreducible cocycle state in 1D.

When $d=3$, our formation unitaries $U(\tau_3)$ correspond to 3-index component tensors $\tritens$, which are harder to characterize. Similar to our $d=2$ proof, color-dependent changes of basis let us rewrite $\tritens$ as a collection of $r$ irreducible tensors, which form the $r$ irreducible $\onethird$-symmetric cocycle states in Theorem~\ref{thm:one}. More precisely, $\tritens$ is irreducible when it cannot be written as the sum of two nonzero tensors with disjoint supports at every index. In $d=3$ however, there is no known analog of Gaussian elimination to efficiently decompose $\tritens$ into irreducible tensors. Nonetheless, we show in Appendix~\ref{sec:thmone} that there is still a normal form letting us prove Theorem~\ref{thm:one} for $d=3$. Consequently, the behavior of general $\onethird$-symmetric cocycle states depends only on the behavior of general irreducible cocycle states.

In the simplest case of $m=1$, the only nontrivial trilinear function is $\omega_3^{(UJ)}$ (defined previously), showing that $\zeta_3(1)=1$. In contrast to the 1D case though, in 2D we find many different irreducible cocycle states, the simplest being shown in Figure~\ref{fig:classification}c. A numerical search shows that $\zeta_3(2)=4$ and $\zeta_3(3)=50$, and we expect infinitely many irreducible states to appear in general $\ztwom$. Despite this difficulty, every irreducible $\onethird$-symmetric cocycle state should clearly contain at least as much usable entanglement as the Union Jack state, which lets us prove a useful operational corollary to Theorem~\ref{thm:one} for $d=3$.

\begin{corollary}
\label{cor:one}
    Let $\tristate$ be a nontrivial $\onethird$-symmetric 3-cocycle state with global symmetry group $\ztwom$ defined on a Union Jack lattice. By appropriate color-dependent changes of basis and non-adaptive single-qubit $Z$ measurements, $\tristate$ can be reduced to $r$ disjoint copies of the Union Jack state, for the same state-dependent $r\geq 1$ as in Theorem~\ref{thm:one}. Consequently, $\tristate$ is a Pauli universal resource state for MQC.
\end{corollary}

We prove Corollary~\ref{cor:one} by showing that any irreducible $\ket{\psi(\gamma_i)}$ is equal in some color-dependent change of generating set to a single copy of the Union Jack state, which is disjoint or ``vertex entangled" with all other virtual qubits. This guarantees that measuring $Z$ on the other virtual qubits leaves only the Union Jack state, up to trivial Pauli byproduct operators. Applying this protocol to each irreducible $\ket{\psi(\gamma_i)}$ in Theorem~\ref{thm:one} then proves Corollary~\ref{cor:one}. Further details are given in Appendix~\ref{sec:corone}.

Having discussed the general classification and computational power of low-dimensional $\onethird$-symmetric cocycle states $\tristate$, we now study their SPTO phases relative to the fractional symmetry group $\Gthird$. This classification relative to $\Gthird$ then determines the SPTO phase of $\tristate$ relative to any subgroup of $\Gthird$, including the usual global symmetry $G$. While $\Gthird \simeq G^3$ as groups, they differ operationally by the former arranging each copy of $G$ on a distinct vertex color (``horizontally"), and the latter arranging each copy on a distinct layer of a single vertex (``vertically"). This allows a simple characterization of the SPTO present in these states (Theorem~\ref{thm:two}).

\begin{theorem}
\label{thm:two}
    Let $\tristate, \ket{\psi(\tau_3')}$ be two $\onethird$-symmetric 2D 3-cocycle states with global symmetry group $G$, where $\tau_3$ and $\tau_3'$ are trilinear functions. If $\tau_3 \neq \tau_3'$, then $\tristate$ and $\ket{\psi(\tau_3')}$ belong to different SPTO phases relative to $\Gthird$. In particular, if $\tau_3$ is nontrivial, then $\tristate$ possesses nontrivial SPTO relative to $\Gthird$.
\end{theorem}

We prove Theorem~\ref{thm:two} by embedding each 3-cocycle state $\tristate$ into a larger Hilbert space associated with $G^3$, where the original $\Gthird$ fractional symmetry of $\tristate$ is simulated using an operationally equivalent $G^3$ global symmetry. This lets us use a known classification of 2D SPTO phases relative to global $G^3$ symmetry to identify each component of $\tritens$ as a unique label of the SPTO phase of $\tristate$, relative to $\Gthird$. Consequently, two states $\tristate, \ket{\psi(\tau_3')}$ are in the same SPTO phase only when their associated tensors $\tritens, \hat{\tau}_3'$ are identical, which proves Theorem~\ref{thm:two}. Further details of our proof are given in Appendix~\ref{sec:thmthree}.

\section{Outlook}

We have shown that computationally universal entanglement is a ubiquitous property of fixed-point states of SPTO with fractional symmetry. While we were able to obtain ``exact" universal resource states in our simple setting of fixed-point model states, more general states with SPTO may require renormalization-style techniques like those of \cite{bartlett2010quantum, miller2015resource, raussendorf2017symmetry, stephen2017determining} to extract their usefulness for MQC, as discussed in more detail in Appendix~\ref{sec:renormalization}. Overall, we expect fractional symmetry to be a powerful tool for guaranteeing certain operational capabilities in more general quantum information processing tasks, such as quantum simulation \cite{cirac2012goals, georgescu2014quantum} and fault-tolerant quantum computation \cite{yoshida2015topological, yoshida2017gapped, roberts2017symmetry}.

\section{Acknowledgements}

This work was supported in part by National Science Foundation grants PHY-1314955, PHY-1521016, and PHY-1620651.

\appendix
\onecolumngrid

\section{Renormalization Group in Measurement-based Quantum Computation}
\label{sec:renormalization}

Our work proves that in the presence of fractional symmetry, the family of 2D $\ztwom$ cocycle states, which describe the renormalization group (RG) fixed-points of SPTO \cite{chen2013symmetry}, are universal resource states for MQC if and only if they possess nontrivial SPTO. Beyond this operational characterization, one might wonder to what extent our results provide evidence that \textit{all} fractionally-symmetric states with nontrivial SPTO are universal resource states for MQC. To properly answer this question, we must address the nature of the RG flow whose fixed points define cocycle states, as well as the relationship of this flow to the RG flow naturally available within the operational setting of MQC. While the relationship between these two flavors of RG flow remains an open question in 2D, in 1D they are known to coincide, providing compelling evidence for the extension of our universality results to characterize entire universal resource phases of SPTO resource states.

The typical form of RG flow used for the study of SPTO is outlined in \cite{chen2013symmetry}, where its associated fixed-point states are shown to be simply cocycle states. This RG flow is implementable by symmetric constant-depth quantum circuits, whose constituent unitary gates are geometrically local in $d-1$ spatial dimensions and each commute with the on-site symmetry group $G$. The use of families of constant-depth unitary circuits for classifying phases of quantum order is well-established in the recent condensed matter literature, and the symmetry-respecting circuits used in this RG flow in particular are guaranteed to preserve the SPTO phase of short-range entangled many-body states. This RG flow lets us transform arbitrarily complex many-body states into simple classes of fixed-point states, allowing for the general study of phases of SPTO resource states in terms of the behavior of a handful of representative examples. Practically speaking, the ability to replicate this RG flow within the operational context of MQC would allow the method of state reduction \cite{chen2010quantum} to be used to promote our proof of the universality of cocycle resource states into a proof that the associated phases of SPTO resource states are universal for MQC.

The RG flow of \cite{chen2013symmetry} is chosen to define fixed-point states which not only have zero correlation length, but which also reflect the algebraic structure inherent in the group cohomological classification of SPTO phases. These states, the cocycle states, can be thought of as the simplest many-body states within a given SPTO phase, whose entire physical behavior is characterized by a single algebraic cocycle. While a ``gauge freedom'' exists between distinct $d$-cocycles in the same cohomology class, we note that a natural choice of gauge is given by the family of $d$-linear cocycles, (at least) one of which exists in each cohomology class \cite{propitius1995topological}. As a result, $d$-linear cocycle states can uniformly be chosen as fixed-points of SPTO, whose properties reflect the general nature of arbitrary many-body states in SPTO phases in $d-1$ spatial dimensions.

From the above, it is clear that the ability to replicate the constant-depth circuit RG flow of \cite{chen2013symmetry} within the operational setting of MQC would allow any resource state with nontrivial 2D $\ztwom$ SPTO to be reduced into a computationally universal trilinear cocycle state, proving all such states to be themselves universal for MQC. While the use of these constant-depth entangling circuits isn't directly permitted within the setting of MQC, an alternate operational notion of RG flow can be achieved using only single-spin measurements. This form of RG flow involves the application of single-spin measurements to certain regions of a many-body resource state, which produces a new renormalized resource state on the unmeasured regions. Although it isn't clear at first sight if single-spin measurements on 2D SPTO resource states are able to reproduce the RG flow needed to map these states to a computationally universal 3-cocycle state, on 1D SPTO resource states these measurements are known to be entirely sufficient. For example, the works of \cite{bartlett2010quantum, miller2015resource} show that patterns of ``$z$-buffering'' measurements can be used to renormalize many-body states in certain nontrivial SPTO phases to fixed-point states whose computational behavior is equivalent to the 1D cluster state. The works of \cite{raussendorf2017symmetry, stephen2017determining} also use RG-style techniques to reproduce the behavior of the 1D cluster state within many-body states belonging to a wide range of 1D SPTO phases. Given these positive results for 1D states, we conjecture that such results should be forthcoming in 2D states, making the search for suitable MQC-compatible RG flows within phases of 2D SPTO a promising research program for finding computationally universal quantum phases of matter.

\section{Group Cohomology and Cocycle States}
\label{sec:cohomology}

In this section, we present a more complete discussion of group cohomology theory and the cocycle state model of \cite{chen2013symmetry}, which sets the stage for our proof of Lemma~\ref{lem:one} in Appendix~\ref{sec:lemone}.

\subsection{Group Cohomology}

Group cohomology theory studies the cohomology groups $\mathcal{H}^{d}(G,U(1))$ associated to a group $G$, for arbitrary $d\geq 0$. For our purposes, the elements of $\mathcal{H}^{d}(G,U(1))$ can be thought of as classifying the SPTO phases of short-range entangled many-body states with global symmetry $G$ in $d-1$ spatial dimensions. The structure of the cohomology groups can be calculated using (inhomogeneous) $d$-cochains, which are arbitrary functions $\xi_d$ mapping $d$-tuples of elements of $G$, $(g_1,g_2,\ldots,g_d) \in G^d$, to individual complex phases $\xi_d(g_1,g_2,\ldots,g_d) \in U(1)$. The set of all $d$-cochains is denoted by $\mathcal{C}^{d}(G,U(1))$, and under pointwise multiplication of function values forms an abelian group isomorphic to $U(1)^{|G|^d}$. The identity element of $\mathcal{C}^{d}(G,U(1))$ is the trivial $d$-cochain $1_d$, which outputs the constant value 1. The ($d$'th) coboundary operator $\partial_d : \mathcal{C}^{d}(G,U(1)) \to \mathcal{C}^{d+1}(G,U(1))$ sends every $d$-cochain $\xi_d$ to a $(d+1)$-cochain $\partial_d\xi_d$, which acts as
\begin{equation}
\label{eq:coboundary}
    \partial_d\xi_d(g_1,\ldots,g_{d+1}) = \xi_d(g_2,\ldots,g_{d+1}) \left\{\prod_{k=1}^d [\xi_d(g_1,\ldots,g_{k-1},g_k g_{k+1},g_{k+2},\ldots,g_{d+1})]^{(-1)^k}\right\}
        [\xi_n(g_1,\ldots,g_d)]^{(-1)^{d+1}}.
\end{equation}

\noindent For example, when $d=2$, $\partial_2\xi_2(g_1,g_2,g_3) = \xi_2(g_2,g_3)\,\xi_2^*(g_1g_2,g_3)\,\xi_2(g_1,g_2g_3)\,\xi_2^*(g_1,g_2)$. The coboundary operator lets us define two important classes of cochains, the cocycles and coboundaries. A $d$-cocycle $\omega_d$ is a $d$-cochain which lies in the kernel of $\partial_d$, satisfying $\partial_d\omega_d = 1_{d+1}$, while a $d$-coboundary $\varphi_d$ is a $d$-cochain which lies in the image of $\partial_{d-1}$, satisfying $\varphi_d = \partial_{d-1}\xi_{d-1}$ for some $(d-1)$-cochain $\xi_{d-1}$. The collection of $d$-cochains and $d$-coboundaries are denoted by $\mathcal{Z}^{d}(G,U(1))$ and $\mathcal{B}^{d}(G,U(1))$ respectively, both of which form subgroups of $\mathcal{C}^{d}(G,U(1))$.

\eqnref{eq:coboundary} can be used to show that $\partial_{d+1}\partial_d\xi_d = 1_{d+2}$ for every $d$-cochain $\xi_d$, which proves the inclusion $\mathcal{B}^{d}(G,U(1)) \subset \mathcal{Z}^{d}(G,U(1))$. The $d$'th cohomology group characterizes the extent to which the reverse inclusion fails to hold, via $\mathcal{H}^{d}(G,U(1)) = \mathcal{Z}^{d}(G,U(1)) / \mathcal{B}^{d}(G,U(1))$. The elements of $\mathcal{H}^{d}(G,U(1))$, the cohomology classes, are represented as equivalence classes of $d$-cocycles modulo $d$-coboundaries, $[\omega_d]_G \in \mathcal{H}^{d}(G,U(1))$, where $[\omega_d']_G = [\omega_d]_G$ if and only if $\omega_d' \, \omega_d\inv = \varphi_d \in \mathcal{B}^{d}(G,U(1))$.

While the above is in principle a complete discussion of the most basic definitions and concepts of group cohomology theory, it is important to recognize that this formalism can also be presented entirely in terms of ``homogeneous" cochains, which are simply reparameterized versions of the inhomogeneous cochains described above. Given an inhomogeneous $d$-cochain $\xi_d(g_1\ldots,g_d)$, we can uniquely define a homogeneous $d$-cochain $\lambda_d(a_0,\ldots,a_d)$, which is related to $\xi_d$ as
\begin{align}
    \lambda_d(a_0,\ldots,a_d) &= \xi_d(a_0\inv a_1,a_1\inv a_2,\ldots,a_{d-1}\inv a_d) \\
    \xi_d(g_1\ldots,g_d) &= \lambda_d(e,g_1,g_1g_2,\ldots,g_1g_2g_3\ldots g_d).
\end{align}

\noindent While homogeneous $d$-cochains are naively functions of $d+1$ arguments, rather than the $d$ arguments appearing in $\xi_d$, this is compensated for by the symmetry $\lambda_d(a_0,\ldots,a_d)=\lambda_d(e,a_0\inv a_1,a_0\inv a_2,\ldots,a_0\inv a_d)$, which holds for all $a_0\in G$. In the setting of homogeneous cochains, the ($d$'th) coboundary operator acts as 
\begin{equation}
\label{eq:coboundary_b}
    \partial_d\lambda_d(a_0,a_1,\ldots,a_{d+1}) = \prod_{k=0}^{d+1} [\lambda_d(a_1,\ldots,a_{k-1},a_{k+1},\ldots,a_{d+1})]^{(-1)^k}.
\end{equation}

\noindent For example, when $d=2$, $\partial_2\lambda_2(a_0,a_1,a_2,a_3) = \lambda_2(a_1,a_2,a_3)\,\lambda_2^*(a_0,a_2,a_3)\,\lambda_2(a_0,a_1,a_3)\,\lambda_2^*(a_0,a_1,a_2)$. Homogeneous $d$-cocycles (resp., $d$-coboundaries) are defined exactly the same as in the inhomogeneous setting, as $d$-cochains lying in the kernel of $\partial_d$ (resp., the image of $\partial_{d-1}$). In what follows, we will denote general homogeneous $d$-cocycles by $\nu_d$.

While we would ideally avoid discussing both forms of $d$-cocycles within the same setting, each form turns out to play a significant role in our work. Inhomogeneous $d$-cocycles serve to capture the algebraic character of group cohomology, and have a close relation to $d$-linear functions, whereas homogeneous $d$-cocycles serve to capture the physical behavior of systems appearing in the $d$-cocycle model. This dual nature is most apparent in Appendix~\ref{sec:lemone}, during our proof of Lemma~\ref{lem:one}, where we start with a homogeneous $\nu_d$ defining a $d$-cocycle state, and end with a proof that the \textit{inhomogeneous} counterpart of $\nu_d$ is a $d$-linear function. This issue is touched upon in more detail in Appendices~\ref{sec:cochain_states} and \ref{sec:lemone}.

\subsection{Cocycle States}
\label{sec:cochain_states}

In this Section, we discuss several important details of the $d$-cocycle state construction \cite{chen2013symmetry}, which we describe here in the more context-appropriate formalism of homogeneous $d$-cocycles. We also discuss a generalization of this construction which outputs states whose SPTO is associated with a lower spatial dimension than that of the defining lattice, and is capable of constructing general stabilizer states \cite{gottesman1997stabilizer}. In addition, we briefly mention the idea of ``$d$-cochain states", a relatively uninteresting generalization of $d$-cocycle states which will be utilized in our proof of Lemma~\ref{lem:one}.

At a broad level, the $d$-cocycle state construction involves converting homogeneous $d$-cocycles $\nu_d$ into diagonal $d$-body unitary gates $U(\nu_d)$, which are applied transversally across a $(d-1)$-dimensional lattice containing $n$ qudits. This prescription outputs a state $\ket{\psi(\nu_d)}$, said to be the $d$-cocycle state generated by $\nu_d$, which we had previously written in inhomogeneous form as $\ket{\psi(\omega_d)}$.

To define our $d$-cocycle state, we need to choose a symmetry $G$, a $d$-cocycle $\nu_d \in \mathcal{Z}^d(G,U(1))$, and a lattice $\Lambda$. We use $G$ to determine the Hilbert space of a single qudit, $H_G$, which is chosen to be the (left) regular representation of $G$. This means that $H_G$ has dimension $|G|$, is spanned by an orthonormal ``$G$ basis" $\{\ket{a}\}_{a\in G}$, and is acted on by $G$ as $X_g\ket{a}=\ket{ga}$, for all $g,a\in G$. $H_G$ contains a unique symmetric state $\ket{+_G}$, given by $\ket{+_G} = \sum_{a\in G}\ket{a}$ (up to normalization).

When $G=\ztwo\simeq\{0,1\}$, $H_G$ corresponds to a single qubit, where the $G$ basis is the usual computational basis and $X_0=I, X_1=X$. For $G=\ztwom$, our main case of interest, we can use the isomorphism $H_{\ztwom} \simeq (H_{\ztwo})^{\otimes m}$ to identify each local spin with a collection of $m$ qubits. This identification depends on our choice of generating set for $\ztwom$, relative to which an arbitrary $G$ basis vector splits as $\ket{a} \simeq \bigotimes_{i=1}^m \ket{a_i}^{(i)}$, with $i$ indexing the $m$ ``virtual" qubits collectively representing $H_{\ztwom}$. However, the identity $|+_{\ztwom}\rangle \simeq \bigotimes_{i=1}^m \ket{+}^{(i)}$ remains true in every generating set, where $\ket{+}:=\frac{1}{\sqrt{2}}(\ket{0}+\ket{1})$.

Given $H_G$, we can use our $d$-cocycle $\nu_d$ to construct a $d$-body ``formation gate" $U(\nu_d)$, which is used to generate $\ket{\psi(\nu_d)}$. This gate is diagonal in the $G$ basis, and has the form of
\begin{align}
\label{eq:cocycle_gate}
    U(\nu_d) &= \sum_{\mathbf{a}\in G^d} \nu_d(e,a^{(1)},a^{(2)},\ldots,a^{(d)}) \ketbra{\mathbf{a}}{\mathbf{a}}.
\end{align}

\noindent Here, $\mathbf{a} = (a^{(1)},\ldots,a^{(d)})$ is a tuple of $d$ group elements $a^{(c)}\in G$ and $\ket{\mathbf{a}} = \bigotimes_{c=1}^d \ket{a^{(c)}}^{(c)}$ is the corresponding $d$-qudit product state, with $c$ indexing the $d$ qudits which $U(\nu_d)$ acts on.

To form our $d$-cocycle state $\ket{\psi(\nu_d)}$, we place a symmetric qudit $\ket{+_G}\in H_G$ at every vertex of our lattice $\Lambda$, then transversally apply $U(\nu_d)$ to the vertices surrounding each $d$-simplex in $\Lambda$. We assume that $\Lambda$ is a $(d-1)$-dimensional, $d$-colorable simplicial complex with closed boundaries and $n$ vertices. In this case, the $d$ vertices surrounding each $d$-simplex $\Delta_d\in \Lambda_d$ are all different colors, which lets us pair the final $d$ indices of $\nu_d$ with the $d$ colors of our lattice in a fixed manner. The transversal application of $U(\nu_d)$ then defines a ``formation circuit" $U_F$, which forms $\ket{\psi(\nu_d)}$ as $\ket{\psi(\nu_d)} = U_F\ket{+_G}^{\otimes n}$. $U_F$ is given by
\begin{equation}
\label{eq:cocycle_state}
    U_F = \prod_{\Delta_d \in \Lambda_d} U(\nu_d)^{s(\Delta_d)},
\end{equation}

\noindent where $U(\nu_d)$ is applied to the $d$ qudits surrounding each $d$-simplex $\Delta_d$, and $s(\Delta_d) = \pm 1$ serves to alternately apply either $U(\nu_d)$ or its complex conjugate. $s(\Delta_d)$ is chosen so that every pair of $d$-simplices $\Delta_d,\Delta_d'$ which overlap along a common $(d-1)$-simplex satisfy $s(\Delta_d) = -s(\Delta_d')$. This alternation turns out to guarantee that $\ket{\psi(\nu_d)}$ is invariant under the global action of $G$ to all $n$ vertices in $\Lambda$, provided $\nu_d$ is a valid $d$-cocycle. Further details on this construction can be found in \cite{chen2013symmetry}.

In addition to the construction introduced above, it is also possible to define cocycle states in $d$ spatial dimensions which are associated with cocycles $\nu_k$, where $k\leq d$. In this variant, the cocycle $\nu_k$ is used to form a $k$-body gate $U(\nu_k)$, which is applied to all $k$-simplices of our lattice in a similar manner as above. The difference here is that these $k$-simplices will be of strictly lower dimension than the dimension of the defining lattice, which gives the associated states different operational properties. For example, the operational classification of these ``lower-dimensional SPTO" cocycle states under finite-depth symmetry-respecting unitaries won't align with the mathematical classification of the cohomology class of $\nu_k$, unless an additional lattice translation symmetry is respected. While this makes such states less interesting from a condensed matter perspective, they still occur frequently within quantum information. In particular, the family of graph states \cite{hein2005entanglement} can be realized (up to local Pauli Z's) as cocycle states associated with the $G=\ztwo$ 2-cocycle $\omega_2^{(1C)}(g,h) = (-1)^{g\cdot h}$, since $U(\omega_2^{(1C)})$ acts on qubits 1 and 2 as the two-qubit gate $U(\omega_2^{(1C)})_{1,2} = Z_1\, (CZ)_{1,2}$. The choice of lattice in this case is the unique graph associated to that graph state. Since all stabilizer states \cite{gottesman1997stabilizer} are equivalent to graph states under local Clifford operations \cite{schlingemann2002stabilizer}, this shows that the family of (extended) cocycle states contains all possible stabilizer states. Further details of this variation of the cocycle state construction can be found in \cite{miller2016hierarchy} and Section XII of \cite{chen2013symmetry}.

It is useful to define another minor generalization of the regular cocycle state construction above, where general $d$-cochains $\lambda_d$ are used to form $(d-1)$-dimensional ``$d$-cochain states" $\ket{\psi(\lambda_d)}$. This construction is essentially identical to the regular $d$-cocycle construction, but with $\lambda_d$ taking the place of $\nu_d$ in \eqnref{eq:cocycle_gate}. The biggest difference between cocycle and cochain states is that the former always possess global $G$ symmetry on closed boundaries, while the latter generally do not. In determining whether a given cochain state $\ket{\psi(\lambda_d)}$ is invariant under a symmetry operation $X_g$, we can always instead determine whether the formation circuit $U_F$ commutes with $X_g$. This is measured by the group commutator, $K(U_F,X_g):=U_F X_g U_F^\dagger X_g^\dagger$. Clearly, $K(U_F,X_g)=I$ implies $X_g\ket{\psi(\lambda_d)}=\ket{\psi(\lambda_d)}$, but $\ket{\psi(\lambda_d)}$ being an even-magnitude superposition of all $G$-basis states means that $X_g\ket{\psi(\lambda_d)}=\ket{\psi(\lambda_d)}$ implies $K(U_F,X_g)=I$ as well.

As a final note, we mention that the concept of ``fractional symmetry'' which we utilize to simplify our study of cocycle and cochain states should not be confused with the phenomena of pairs of linked projective representations of a group $G$, which is sometimes referred to as ``fractional symmetry'' in certain condensed matter settings.

\section{Proof of Lemma~\ref{lem:one}}
\label{sec:lemone}

This Appendix is intended to give the complete proof of our Lemma~\ref{lem:one}, which states that every $\oned$-symmetric $d$-cocycle state is generated by a unique $d$-linear function $\tau_d$, for $d\leq 3$. While the $d=1$ case is trivial (1-cochains are the same as 1-linear functions), in Sections~\ref{sec:lemonetwo} and \ref{sec:lemonefour} we give proofs of Lemma~\ref{lem:one} for $d=2$ and $d=3$, respectively. These proofs each rely upon certain technical results, which guarantee that any $(d-1)$-cochain state (see Section~\ref{sec:cochain_states}) with global symmetry is generated by a $(d-1)$-cocycle, up to boundary terms. These results are given for $d=2$ and $d=3$ in Sections~\ref{sec:lemoneone} and \ref{sec:lemonethree}, respectively. While we expect that this process can be inductively continued to give a proof of Lemma~\ref{lem:one} for arbitrary $d$, we focus here on developing a physically motivated proof applicable to our computationally relevant setting of 1D and 2D MQC resource states.

In the following, we refer in several places to ``standard $d$-cocycle identities" (for $d=2$ or 3), which we typically mean to be rearrangements/reparameterizations of one of the cocycle relations
\begin{align}
    \nu_2(e,a,b)\nu_2^*(e,g\inv a,g\inv b) &= \nu_2(e,g,b)\nu_2^*(e,g,a) \\
    \nu_3(e,a,b,c)\nu_3^*(e,g\inv a,g\inv b,g\inv c) &= \nu_3(e,g,b,c)\nu_3^*(e,g,a,c)\nu_3(e,g,a,b). 
\end{align}

We will frequently use parameterized cochains or cocycles, which will be written with the variable parameter separated from the regular function arguments by a semicolon. For example, a term of the form $\lambda_2(g;a,b)$ will indicate a $g$-parameterized family of homogeneous 2-cochains with inputs $a$ and $b$. When $U_F$ and $X_g$ are unitary operators, we will use $K(U_F,X_g):=U_FX_gU_F^\dagger X_g^\dagger$ to indicate their group commutator (as in Section~\ref{sec:cochain_states}), but we will also use this notation to indicate an appropriately reparameterized product of cochains when the first argument is replaced by a $d$-cochain $\lambda_d$. For example, the expression $K(\nu_1(e,a),X_g)=\nu_1(e,a)X_g\nu_1^\dagger(e,a)X_g^\dagger$ is simply $\nu_1(a)\nu_1^\dagger(g\inv a)$. Our indexing of physical sites in the following will generally be limited to local regions whose sites are labeled as A, B, B',$\ldots$, and whose corresponding group elements are labeled as $a,b,b'$, etc. Symmetry operators are labeled analogously, and we will use $X_g^{(A,B)}$ to indicate a symmetry operator acting on sites $A$ and $B$.

\subsection{Symmetric 1-cochain states are generated by 1-cocycles}
\label{sec:lemoneone}

We first prove that any 1-cochain state $\ket{\psi(\lambda_1)}$ defined on closed boundaries which is invariant under global $G$ symmetry is equivalently a 1-cocycle state, and that its 1-cochain $\lambda_1$ is a 1-cocycle up to overall phase. This result will be used in Section~\ref{sec:lemonetwo} to prove the $d=2$ version of Lemma~\ref{lem:one}.

\begin{figure}[t]
  \centering
  \includegraphics[width=0.9\textwidth]{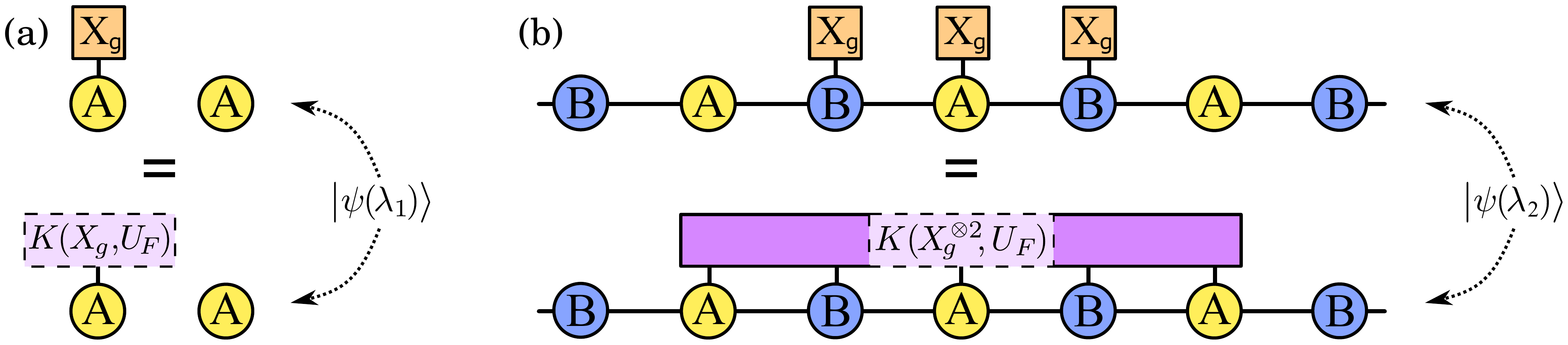}
  \caption{Reduction of global symmetry conditions for $d$-cochain states on closed boundaries $\ket{\psi(\lambda_d)}$ to local algebraic conditions for $d$-cochains $\lambda_d$, where $d=1,2$. (a) 1-cochain states on closed boundaries in 0D are simply pairs of unentangled spins. For $\ket{\psi(\lambda_1)}$ to possess global symmetry, the commutator of the formation circuit $U_F$ with $X_g$ applied to one spin must be constant at that spin, and consequently must be a complex phase. (b) For a 1D 2-cochain state $\ket{\psi(\lambda_1)}$ to possess global symmetry, the commutator of the formation circuit $U_F$ with $X_g$ applied to a region of our system must be constant within the interior of that region (central A site, light purple), but is generally nontrivial near the region's boundaries (dark purple).}
  \label{fig:lemma_symmetry}
\end{figure}

For our state to have ``closed boundaries" in 0D, we can wlog choose our global Hilbert space to consist of two separated spins, as shown in Figure~\ref{fig:lemma_symmetry}a. Our global formation circuit is then $U_F = \lambda_1(e,a) \lambda_1^\dagger(e,a')$, and the condition for global symmetry (equivalently, $\frac{1}{1}$-symmetry) is that $K((X_g^{\otimes2})^\dagger,U_F) = I$. This means,
\begin{align}
    K((X_g^{\otimes2})^\dagger,U_F) &= \left(\lambda_1(e,ga)\lambda_1^*(e,a)\right) \left(\lambda_1(e,ga')\lambda_1^*(e,a')\right)^* = 1. \label{eq:3}
\end{align}

\noindent It is helpful here to define $\omega_1(g;a):= \lambda_1(e,ga)\lambda_1^*(e,a)$. In this case, \eqnref{eq:3} requires $\omega_1(g;a) = \omega_1(g;e)$, so that $\omega_1(g;a)$ is independent of $a$. This lets us abbreviate $\omega_1(g):=\omega_1(g;e)$, in which case $\lambda_1(e,a) = \omega_1(a)\lambda_1(e,e) = \alpha\,\omega_1(a)$, where $\alpha:=\lambda_1(e,e)$ contributes only a constant overall phase.

Using the above relations, we find that
\begin{align}
    \omega_1(gh) &= \lambda_1(e,gh)\lambda_1^*(e,e) = \left(\lambda_1(e,gh)\lambda_1^*(e,h)\right) \left(\lambda_1(e,h)\lambda_1^*(e,e)\right) = \omega_1(g;h)\,\omega_1(h;e) \\
    &= \omega_1(g)\,\omega_1(h).
\end{align}

\noindent This is equivalent to $\partial_1\omega_1(g,h) = \omega_1(h)\omega_1^*(gh)\omega_1(g) = 1$, which proves that $\omega_1$ is a valid 1-cocycle. This shows that any symmetric 1-cochain state is generated by a unique 1-cocycle $\omega_1$, and that the associated 1-cochain is proportional to that 1-cocycle as $\lambda_1(e,a) = \alpha\,\omega_1(a)$.

\subsection{$\onehalf$-symmetric 2-cocycle states are generated by bilinear functions (Lemma~\ref{lem:one}, $\mathbf{d=2}$)}
\label{sec:lemonetwo}

We now move to the case of 1D 2-cocycle states with $\onehalf$-symmetry, which we will show are each generated by a unique bilinear function. Furthermore, we will show that the associated 2-cocycle itself must be a bilinear function up to overall phase. Although $\onehalf$-symmetry is a global condition on our state $\ket{\psi(\nu_2)}$, it is simple to reduce this to a local condition. In particular, if our global formation circuit is $U_F$, then we must have the commutator of $U_F$ with an A-site symmetry operator, $K(U_F,X_g^{(A)})$, be trivial at this A site. This commutator is given by 
\begin{align}
    K(U_F, X_g^{(A)}) &= \left(\nu_2(e,a,b) \nu_2^*(e,a,b')\right) \left(\nu_2(e,g\inv a,b) \nu_2^*(e,g\inv a,b')\right)^* \\
    &= \nu_2(g\inv a,a,b) \nu_2^*(g\inv a,a,b') = \lambda_1(g;a,b) \lambda_1^*(g;a,b'),
\end{align}

\noindent where $\lambda_1(g;a,b):=\nu_2(g\inv a,a,b)$, and the second equality comes from a standard 2-cocycle identity. It is clear that $\lambda_1(g;a,b)$ is a 1-cocycle with respect to $a$ and $b$, and the above commutator being trivial on its associated A site is equivalent to $\lambda_1(g;a,b)$ generating a symmetric 0D state on closed boundaries. From Section~\ref{sec:lemoneone}, we know that $\lambda_1(g)$ must have the form $\lambda_1(g;a,b)=\alpha(g)\,\omega_1(g;a\inv b)$, where each $\omega_1(g)$ is a 1-cocycle of $G$. Consequently, $\nu_2(e,a,b)=\alpha(a)\,\omega_1(a;a\inv b)$.

We now wish to use the 2-cocycle nature of $\nu_2$ to constrain the manner in which the phases $\alpha(a)$ and 1-cocycles $\omega_1(a;a\inv b)$ vary with $a$. If we take the commutator of a single $\nu_2(e,a,b)$ with the global symmetry operator $X_g^{\otimes n}$, then we find
\begin{align}
    K(\nu_2(e,a,b), X_g^{\otimes n}) &= \nu_2(e,a,b) \nu_2^*(e,g\inv a,g\inv b) = \alpha(a)\alpha^*(g\inv a) \omega_1(a;a\inv b)\omega_1^*(g\inv a;a\inv b) \label{eq:two_coc_a} \\
    &= \nu_2^*(e,g,a) \nu_2(e,g,b) = \omega_1(g;a\inv b). \label{eq:two_coc_b}
\end{align}

\noindent In \eqnref{eq:two_coc_a}, we directly substitute $\alpha(a)\,\omega_1(a;a\inv b)$ for each factor of $\nu_2(e,a,b)$, whereas in \eqnref{eq:two_coc_b} we first use a standard 2-cocycle identity, then substitute for each $\nu_2$ term. In order for these expressions to be equivalent, we must have the following equality hold for all $g,h,f\in G$:
\begin{equation}
    \omega_1^*(gh;f)\omega_1(g;f)\omega_1(h;f) = \alpha(gh)\alpha^*(h). \label{eq:two_coc_constraint} \\
\end{equation}

\noindent \eqnref{eq:two_coc_constraint} allows us to determine how $\alpha(g)$ and $\omega_1(g;f)$ depend on their first arguments. In particular, setting $g=e$ in \eqnref{eq:two_coc_constraint} shows that $\omega_1(e;f)=1$ for all $f$, while setting $h=e$ shows that $\alpha(g)=\alpha(e)$ for all $g$. Consequently, \eqnref{eq:two_coc_constraint} reduces to $\omega_1(gh;f) = \omega_1(g;f)\omega_1(h;f)$, so that $\omega_1(g;f)$ acts as a unitary character of $G$ in its first argument. As $\omega_1(g;f)$ was already chosen to be a unitary character of $G$ in its second argument, this shows that $\tau_2(g,f):=\omega_1(g;f)$ is a bilinear function of $G$. Since $\nu_2(e,a,b)=\alpha(a)\omega_1(a;a\inv b)=\alpha\,\tau_2(a,a\inv b)$ (with $\alpha:=\alpha(e)$), this completes our proof that every $\onehalf$-symmetric 2-cocycle state is generated by a unique bilinear function $\tau_2$, and that the inhomogeneous form of the associated 2-cocycle $\nu_2$ is proportional to $\tau_2$.

\subsection{Symmetric 2-cochain states are generated by 2-cocycles}
\label{sec:lemonethree}

As a 1D generalization of our proof in Section~\ref{sec:lemoneone}, we will find a necessary and sufficient algebraic condition which a 2-cochain must satisfy to define a symmetric 2-cochain state on closed boundaries. We will show that any such 2-cochain state is generated by a unique 2-cocycle $\nu_2$, and that its associated 2-cochain can be factorized as $\lambda_2(e,a,b) = \nu_2(e,a,b) \lambda_1^{(A)*}(e,a)\lambda_1^{(B)}(e,b)$. Here, $\lambda_1^{(A)}$ and $\lambda_1^{(B)}$ are homogeneous 1-cochains which end up only acting on the boundaries of our system. Consequently, the state generated by $\lambda_2$ on closed boundaries is exactly the same as that generated by $\nu_2$, so that we can always associated a unique 2-cocycle to every globally symmetric 2-cochain state. This result will be used in Section~\ref{sec:lemonefour} to prove the $d=3$ version of Lemma~\ref{lem:one}.

The condition of being symmetric on closed boundaries in 1D is that the commutator $K(X_{g\inv}^{\otimes n},U_F)$ of our formation circuit with arbitrary global symmetry operators is globally trivial. At a local level, this commutator can be expressed as a product of two-body ``defect gates" associated with the commutator $\eta_2(g;e,a,b):=K(X_g^{(A,B)\dagger},\lambda_2(e,a,b))$, and global symmetry requires the product of these defect gates to be trivial in the interior of any region it is being transversally applied to (see Figure~\ref{fig:lemma_symmetry}b). This implies that the A-interior product $\eta_2(g;e,a,b)\eta_2^*(g;e,a,b')$ should be independent of $a$, and that the B-interior product $\eta_2(g;e,a,b)\eta_2^*(g;e,a',b)$ should be independent of $b$. As we show below, these two algebraic conditions alone are sufficient to show that the state generated by $\lambda_2$ is a 2-cocycle state.

Before going further, we first derive an important consistency relation involving $\eta_2$. Since $\eta_2(g;e,a,b)=\lambda_2(e,ga,gb)\lambda_2^*(e,a,b)$, we have that
\begin{align}
    \eta_2(gh;e,a,b) &= \lambda_2(e,gha,ghb)\lambda_2^*(e,a,b) \\
    &= \left(\lambda_2(e,gha,ghb)\lambda_2^*(e,ha,hb)\right) \left(\lambda_2(e,ha,hb)\lambda_2^*(e,a,b)\right) \\
    &= \eta_2(g;e,ha,hb) \eta_2(h;e,a,b). \label{eq:defect_consistency}
\end{align}

\noindent We will make use of \eqnref{eq:defect_consistency} shortly, but for now focus on expanding the algebraic conditions mentioned above which arise from global symmetry. These conditions require that certain overlapping products of $\eta_2(g;e,a,b)$ terms must be independent of the value taken within their region of overlap. This can be expressed as the vanishing of double commutators
\begin{align}
    K(X_h^{(A)\dagger},\eta_2(g;e,a,b)\eta_2^*(g;e,a,b')) &= \left(\eta_2(g;e,ha,b)\eta_2^*(g;e,a,b)\right) \left(\eta_2(g;e,ha,b')\eta_2^*(g;e,a,b')\right)^* = 1 \label{eq:ddefect_a} \\
    K(X_h^{(B)\dagger},\eta_2(g;e,a,b)\eta_2^*(g;e,a',b)) &= \left(\eta_2(g;e,a,hb)\eta_2^*(g;e,a,b)\right) \left(\eta_2(g;e,a',hb)\eta_2^*(g;e,a',b)\right)^* = 1 \label{eq:ddefect_b}
\end{align}

\noindent In order for Eqn.'s~\ref{eq:ddefect_a} and \ref{eq:ddefect_a} to hold, we must have $\eta_2(g;e,ha,b)\eta_2^*(g;e,a,b)$ be independent of $b$, and $\eta_2(g;e,a,hb)\eta_2^*(g;e,a,b)$ be independent of $a$, which lets us define $\omega_2^{(A)}(g,h;e,a):=\eta_2(g;e,ha,e)\eta_2^*(g;e,a,e)$ and $\omega_2^{(B)}(g,h;e,b):=\eta_2(g;e,e,hb)\eta_2^*(g;e,b)$. This allows us to express $\eta_2(g;e,a,b)$ as
\begin{align}
    \eta_2(g;e,a,b) &= \omega_2^{(A)}(g,a;e,e) \eta_2(g;e,e,b) = \alpha(g)\, \omega_2^{(A)}(g,a) \omega_2^{(B)}(g,b), \label{eq:defect_structure}
\end{align}

\noindent where $\alpha(g):=\eta_2(g;e,e,e)$, and we have chosen to abbreviate $\omega_2^{(A)}(g,a):=\omega_2^{(A)}(g,a;e,e)$ and $\omega_2^{(B)}(g,b):=\omega_2^{(B)}(g,b;e,e)$. We can now insert this expression for $\eta_2$ into \eqnref{eq:defect_consistency}, which gives the following consistency relation between $\alpha$, $\omega_2^{(A)}$, and $\omega_2^{(B)}$
\begin{align}
    \partial_1\alpha(g,h) &= \alpha(g) \alpha^*(gh) \alpha(h) \nonumber \\
    &= \left(\omega_2^{(A)}(g,ha)\omega_2^{(A)*}(gh,a)\omega_2^{(A)}(h,a)\right)^* \left(\omega_2^{(B)}(g,hb)\omega_2^{(B)*}(gh,b)\omega_2^{(B)}(h,b)\right)^* \label{eq:defect_consistency_b} \\
    &= \omega_2^{(A)*}(g,h) \omega_2^{(B)}(g,h), \nonumber
\end{align}

\noindent where the last equality in \eqnref{eq:defect_consistency_b} comes from setting $a=b=e$ and using the fact that $\omega_2^{(A)}(h,e)=\omega_2^{(B)}(h,e)=1$. This is allowed, since the second equality in \eqnref{eq:defect_consistency_b} reveals this quantity to be independent of $a$ and $b$. Consequently, we have that $\omega_2^{(A)}(g,a)=\left(\omega_2^{(B)}(g,a) \,\partial_1\alpha(g,a)\right)^*$, which can be used to revise our expression for $\eta_2(g;e,a,b)=\alpha(ga)\alpha^*(a)\omega_2^{(B)*}(g,a)\omega_2^{(B)}(g,b)$. We insert this back into \eqnref{eq:defect_consistency} to obtain
\begin{equation}
    \left(\omega_2^{(B)}(g,ha)\omega_2^{(B)*}(gh,a)\omega_2^{(B)}(h,a)\right)\left(\omega_2^{(B)}(g,hb)\omega_2^{(B)*}(gh,b)\omega_2^{(B)}(h,b)\right)^* = 1. \label{eq:defect_consistency_c}
\end{equation}

\noindent Because of the factorization of \eqnref{eq:defect_consistency_c} into terms which depend only on $a$ or only on $b$, each term must be equal to some function of $g$ and $h$ alone, which we will call $\phi(g,h):=\omega_2^{(B)}(g,hb)\omega_2^{(B)*}(gh,b)\omega_2^{(B)}(h,b)$. Setting $b=e$ in this expression for $\phi(g,h)$ reveals that $\phi(g,h)=\omega_2^{(B)}(g,h)$, which is a key result. We can insert this into our definition of $\phi(g,h)$ to show that $\omega_2^{(B)}$ is a valid inhomogeneous 2-cocycle, via
\begin{equation}
    \partial_2\omega_2^{(B)}(g,h,b) = \omega_2^{(B)}(h,b)\,\omega_2^{(B)*}(gh,b)\,\omega_2^{(B)}(g,hb)\,\omega_2^{(B)^*}(g,h) = 1.
\end{equation}

\noindent With this 2-cocycle in hand, we can quickly express our original 2-cochain $\lambda_2$ as a homogeneous 2-cocycle with additional boundary terms. In particular, from our original definition of $\eta_2(g;e,a,b)$, we see that $\lambda_2(e,a,b) = \eta_2(a;e,e,a\inv b)\,\lambda_1(a,b)$, where $\lambda_1(a,b):=\lambda_2(e,e,a\inv b)$ is a homogeneous 1-cochain. We can use this to obtain
\begin{align}
    \lambda_2(e,a,b) &= \alpha^*(e)\omega_2^{(B)*}(a,e)\,\alpha(a)\lambda_1(a,b)\,\omega_2^{(B)}(a,a\inv b) \\
    &= \left(\partial_1\lambda_1(e,a,b)\,\omega_2^{(B)}(a,a\inv b)\right)\left(\alpha(a)\lambda_1^*(e,a)\right)\lambda_1(e,b) \\
    &= \nu_2(e,a,b)\lambda_1^{(A)*}(e,a)\lambda_1^{(B)}(e,b). \label{eq:cochain_structure}
\end{align}

\noindent In the second equality above, we have chosen to rewrite $\lambda_1(a,b)$ in terms of the homogeneous 2-coboundary $\partial_1\!\lambda_1(e,a,b)$, whereas in the third equality, we have defined $\lambda_1^{(A)}(e,a):=\alpha^*(a)\lambda_1(e,a)$, $\lambda_1^{(B)}(e,b):=\lambda_1(e,b)$, and $\nu_2(e,a,b):=\partial_1\!\lambda_1(e,a,b)\,\omega_2^{(B)}(a,a\inv b)$. From its definition, it is clear that $\nu_2(e,a,b)$ is a valid homogeneous 2-cocycle, and this consequently completes our proof that any symmetric 2-cochain state is generated by a unique 2-cocycle.

\subsection{$\onethird$-symmetric 3-cocycle states are generated by trilinear functions (Lemma~\ref{lem:one}, $\mathbf{d=3}$)}
\label{sec:lemonefour}

We now consider 2D 3-cocycle states with $\onethird$-symmetry, which we will show are each generated by a unique trilinear function. Furthermore, we will show that the associated 3-cocycle must itself be a trilinear function, up to terms which only act on the boundaries of our system. We take a similar approach as was done in Section~\ref{sec:lemonetwo}, where we reduce the global $\onethird$-symmetry condition on the state $\ket{\psi(\nu_3)}$ to a local algebraic condition on the 3-cocycle $\nu_3$. If our global formation circuit is $U_F$, then we must have the commutator of $U_F$ with an A-site symmetry operator, $K(U_F,X_g^{(A)})$, be trivial at this A site. This commutator, which is only supported on the B and C sites surrounding our A site, is given by
\begin{align}
    K(U_F, X_g^{(A)}) &= \prod_{<i,j>}\left(\nu_3(e,a,b_i,c_j) \nu_3^*(e,g\inv a,b_i,c_j)\right) \\
    &= \prod_{<i,j>}\nu_3(g\inv a,a,b_i,c_j) = \prod_{<i,j>}\lambda_2(g;a,b_i,c_j),
\end{align}

\noindent where $\lambda_2(g;a,b_i,c_j):=\nu_3(g\inv a,a,b_i,c_j)$, and the second equality comes from using a standard 3-cocycle identity. It is clear that $\lambda_2(g;a,b,c)$ is a 2-cocycle with respect to $a$, $b$, and $c$, and the above commutator being trivial on its associated A site is equivalent to $\lambda_2(g;a,b,c)$ generating a symmetric 1D state on closed boundaries. From Section~\ref{sec:lemonethree}, we know that each $\lambda_2(g)$ must have the form $\lambda_2(g;a,b,c)=\nu_2(g;a,b,c)\lambda_1^{(B)*}(g;a,b)\lambda_1^{(C)}(g;a,c)$, where each $\nu_2(g)$ is a 2-cocycle, and all $\lambda_1^{(B)}(g)$'s, $\lambda_1^{(C)}(g)$'s are 1-cochains. While this condition on $\lambda_2(g)$ comes only from assuming fractional symmetry with respect to A-site symmetries, full $\onethird$-symmetry requires more, namely that each $\lambda_2(g)$ generates a $\onehalf$-symmetric 1D state. From Section~\ref{sec:lemonetwo}, we know that this forces each $\nu_2(g)$ to be associated with a unique bilinear function $\tau_2(g)$. Consequently, we can make use of the ansatz $\nu_3(e,a,b,c) = \lambda_2(a;a,b,c) = \tau_2(a;a\inv b,b\inv c)\lambda_1^*(a;a,b)\lambda_1(a;a,c)\beta_1(a;a,b)$, where $\lambda_1(g;a,c):=\lambda_1^{(C)}(g;a,c)$ and $\beta_1(g;a,b):=\lambda_1^{(B)*}(g;a,b)\lambda_1^{(C)}(g;a,b)$.

With this ansatz for $\nu_3$ in hand, we can now use the 3-cocycle nature of $\nu_3$ to constrain the manner in which the 1-cochains $\lambda_1(g;a,c)$ and $\beta_1(g;a,b)$, as well as the bilinear 2-cocycles $\tau_2(g;a,b,c)$, vary with $g$. If we take the commutator of $\nu_3(e,a,b,c)$ with the global symmetry operator $X_g^{\otimes n}$, we obtain
\begin{align}
    K(\nu_3(e,a,b,c),X_g^{\otimes n}) &= \nu_3(e,a,b,c) \nu_3^*(e,g\inv a,g\inv b,g\inv c) \\
    &= \tau_2(a;a\inv b,b\inv c)\tau_2^*(g\inv a;a\inv b,b\inv c) \times\! \left(\frac{\lambda_1^*(a;a,b)\lambda_1(a;a,c)\beta_1(a;a,b)}{\lambda_1^*(g\inv a;a,b)\lambda_1(g\inv a;a,c)\beta_1(g\inv a;a,b)}\right) \label{eq:three_coc_a} \\
    &= \nu_3(e,g,a,b)\nu_3^*(e,g,a,c)\nu_3(e,g,b,c) = \tau_2(g;a\inv b,b\inv c) \beta_1(g;g,b). \label{eq:three_coc_b}
\end{align}

\noindent In \eqnref{eq:three_coc_a}, we directly substitute our ansatz for each factor of $\nu_3$, whereas in \eqnref{eq:three_coc_b} we first use a standard 3-cocycle identity, then substitute for each $\nu_3$ factor. In order for these expressions to be equal, we must have the following hold for all $g,h,f\in G$:
\begin{equation}
    \tau_2^*(a;a\inv b,b\inv c)\tau_2(g;a\inv b,b\inv c)\tau_2(g\inv a;a\inv b,b\inv c) = \frac{\lambda_1^*(a;a,b)\lambda_1(a;a,c)\beta_1(a;a,b)\beta_1^*(g;g,b)}{\lambda_1^*(g\inv a;a,b)\lambda_1(g\inv a;a,c)\beta_1(g\inv a;a,b)}. \label{eq:three_coc_constraint_a}
\end{equation}

\noindent \eqnref{eq:three_coc_constraint_a} can be used to generate general constraints on the $\lambda_1(g;a,c)$, $\beta_1(g;a,b)$, and $\tau_2(g;a,b,c)$ by considering particular values of $g$, $a$, $b$, and $c$. To begin with, setting $g=e$ gives the constraint $\tau_2(e;a\inv b,b\inv c)=\beta_1^*(e;e,b)$, which holds for all values of $a$, $b$, and $c$. Choosing $a=b=c$ shows that $\beta_1(e;e,a)=\tau_2(e;e,e)=1$, where the last equality comes from the fact that $\tau_2(g;h,f)=1$ whenever $h=e$ or $f=e$. Consequently, this shows that $\beta_1(e;a,b)=\tau_2(e;a\inv b,b\inv c)=1$.

We can now set $b=c$ to obtain $\beta_1(a;a,b)\beta_1^*(g\inv a;a,b)\beta_1^*(g;g,b)=1$, which shows that the factors of $\beta_1$ appearing in \eqnref{eq:three_coc_constraint_a} are collectively trivial. This lets us update \eqnref{eq:three_coc_constraint_a} to read
\begin{equation}
    \tau_2^*(a;a\inv b,b\inv c)\tau_2(g;a\inv b,b\inv c)\tau_2(g\inv a;a\inv b,b\inv c) = \frac{\lambda_1^*(a;a,b)\lambda_1(a;a,c)}{\lambda_1^*(g\inv a;a,b)\lambda_1(g\inv a;a,c)}. \label{eq:three_coc_constraint_b}
\end{equation}

\noindent We can now set $a=b$ in the above expression to obtain $\lambda_1(a;a,c)\lambda_1^*(g\inv a;a,c)=\delta(a)\delta_1^*(g\inv a)$, where we have defined $\delta_1(a):=\lambda_1(a;e,e)$. This substitution allows us to replace the entire right hand side of \eqnref{eq:three_coc_constraint_b}, obtaining
\begin{equation}
    \tau_2^*(a;a\inv b,b\inv c)\tau_2(g;a\inv b,b\inv c)\tau_2(g\inv a;a\inv b,b\inv c) = 1. \label{eq:three_coc_constraint_c}
\end{equation}

\noindent \eqnref{eq:three_coc_constraint_c} clearly shows that the three-index function $\tau_3(g,h,f):=\tau_2(g;h,f)$ is a unitary character of $G$ with respect to its first argument, while the bilinear nature of each $\tau_2(g)$ guarantees that $\tau_3$ is a character in its other arguments as well. Consequently, this proves that $\tau_3$ is a trilinear function of $G$, and that our original 3-cocycle is related to $\tau_3$ as $\nu_3(e,a,b,c) = \tau_3(a,a\inv b,b\inv c)\lambda_1^{(B)*}(a;a,b)\lambda_1^{(C)}(a;a,c)$. Since these last two parameterized 1-cochains act as 2-body terms between A and B (or A and C) sites, they multiply away on closed boundaries, with the resulting state identical to that generated by $\tau_3$, i.e. $\ket{\psi(\nu_3)}=\tristate$. This completes our proof of Lemma~\ref{lem:one} for $d=3$.

\section{Proofs of Theorem~\ref{thm:one} ($\mathbf{d=3}$), Corollary~\ref{cor:one}, and Theorem~\ref{thm:two}}
\label{sec:misc_proofs}

In this Appendix, we will describe the proofs of Theorem~\ref{thm:one} (for $d=3$), Corollary~\ref{cor:one}, and Theorem~\ref{thm:two}, all of which involve a study of the 3-index component tensors $\tritens$ associated with general $\onethird$-symmetric 3-cocycle states. Theorem~\ref{thm:one} is proved as a straightforward consequence of our choice of disjoint normal form for trilinear functions. Corollary~\ref{cor:one} is proved by finding a simple change of basis which transforms $\tritens$ into an ``edge disjoint" normal form, which is compatible with the normal form of Theorem~\ref{thm:one}. Theorem~\ref{thm:two} is proved using simple operational arguments, and some known results about the structure of 2D SPTO phases relative to abelian $G$.

Before we begin these proofs, let's briefly review the role of the 3-index binary tensor $\tritens$ in structuring our 3-cocycle state $\tristate$. As we know from Lemma~\ref{lem:one}, any $\onethird$-symmetric 3-cocycle state $\ket{\psi(\omega_3)}$ on closed boundaries is generated by a unique trilinear function $\tau_3$, so that $\ket{\psi(\omega_3)} = \tristate$. When $G=\ztwom$, this lets us drastically simplify our description of $\ket{\psi(\omega_3)}$, from the $\mathcal{O}(|G|^2)=\mathcal{O}(4^{m})$ complex parameters needed to describe $\omega_3$ to the $m^3$ binary components describing $\tau_3$. By fixing a generating set, we can naturally arrange these components in an $m\times m\times m$ binary tensor $\tritens$, whose algebraic properties encode details of the many-body entanglement structure of $\tristate$. For example, we showed previously that for $d=2$, the rank $r$ of the component matrix $\bitens$ is equal to the (logarithm of the) Schmidt rank of our state $\ket{\psi(\tau_2)}$ across any bipartition, owing to $\ket{\psi(\tau_2)}\simeq\ket{\psi_{1C}}^{\otimes r}$.

When a fixed generating set is used, each tensor $\tritens$ uniquely describes the many-body state $\tristate$; however, the lack of a canonical choice of generating set means that two states with non-identical tensors $\tritens$ and $\tritens'$ might nonetheless be related by a local change of basis. To remedy this issue, we can analyze the ``local unitary orbits" of $\tritens$ under index-dependent changes of basis, which act on each of $\tritens$'s indices as invertible matrices over $GF(2)$, $\chi_A$, $\chi_B$, and $\chi_C$. Using a decomposition of $\chi_A,\chi_B,\chi_C$ into elementary matrices (those which implement elementary row and column operations), it is easy to show that these changes of basis manifest physically as a product of Clifford $CNOT$ gates on the $m$ virtual qubits at each site, and consequently have no influence on the Pauli universality of our $\tristate$.

\subsection{Theorem~\ref{thm:one} ($\mathbf{d=3}$)}
\label{sec:thmone}

Our proof of Theorem~\ref{thm:one} for $d=3$ involves a basic use of the normal form we have chosen for our 3-index tensor $\tritens$. We first define the support of $\tritens$ at a specific index, say C, which is a certain subgroup $S_C\subseteq\ztwom$ associated to $\tau_3$. Our definition of the support is slightly unusual, as the collection of C-site unitary characters $\omega_1^{(a_0,b_0)}(c)$ obtained by fixing arbitrary A- and B-site arguments. Mathematically, $S_C\simeq\{\omega_1^{(a_0,b_0)}(c):=\tau_3(a_0,b_0,c)\,|\,(a_0,b_0)\in G^2\}$. An obvious generalization of this definition is used for the A- and B-site supports $S_A$ and $S_B$ of $\tritens$.

We choose our disjoint normal form to capture the maximal possible decomposition of $\tritens$ into tensors $\tritens^{(i)}$ with completely disjoint supports. Mathematically, this means that if $\tritens=\sum_{i=1}^r\tritens^{(i)}$, and if $S_A^{(i)}$, $S_B^{(i)}$, and $S_C^{(i)}$ indicate the index-specific supports of $\tritens^{(i)}$, then $S_A^{(i)}\cap S_A^{(j)}=S_B^{(i)}\cap S_B^{(j)}=S_C^{(i)}\cap S_C^{(j)}=\emptyset$ for all $i\neq j$. If $\tritens$ is nontrivial and there is no nontrivial decomposition of $\tritens$ into multiple disjoint tensors, then we say that $\tritens$ is irreducible. When $\tritens$ is the sum of $r$ disjoint irreducible tensors, $\tristate$ is clearly the tensor product of $r$ irreducible 3-cocycle states. Consequently, proving that this normal form exists and is well-defined proves the physical decomposition stated in Theorem~\ref{thm:one}.

The existence of our disjoint normal form for all tensors $\tritens$ can be proved through a simple argument. $\tritens$ can be graphically represented as in Figure~2, as a collection of triangles arranged between $3m$ vertices. If these triangles can be grouped into two sets with no mutual intersections (triangles from one set don't intersect those from the other), then we immediately obtain a decomposition of $\tritens$ into two disjoint tensors. Conversely, when $\tritens$ is irreducible, an exhaustive search of all possible index-dependent changes of basis will find $\tritens$ to never have a representation as two sets of disjoint triangles. Performing this search will therefore end in either a decomposition of $\tritens$ into two disjoint tensors, or a proof that $\tritens$ is irreducible. Continuing this inductively will eventually obtain our desired disjoint normal form, which completes our proof of Theorem~\ref{thm:one} for $d=3$.

Because of its use of exhaustive search, the runtime of our tensor normal form algorithm is clearly exponential in $m$. This should be contrasted to the proof of Theorem~\ref{thm:one} when $d=2$, where the normal form of $\bitens$ was obtained efficiently using Gaussian elimination. While a similarly efficient algorithm is obviously desirable here, we emphasize that our above algorithm is still sufficient to prove that $\tristate$ is isomorphic to a product of $r$ mutually unentangled irreducible 3-cocycle states, with a unique $r$. Furthermore, we will see in Section~\ref{sec:corone} that any nontrivial $\tristate$ can still be efficiently utilized as a Pauli-universal resource state, without any knowledge of its decomposition into irreducible states. Consequently, the existence or nonexistence of an efficient algorithm to compute this normal form has no impact on our ability to utilize $\onethird$-symmetric cocycle states as Pauli-universal resource states.

\subsection{Corollary~\ref{cor:one}}
\label{sec:corone}

Here we give a proof of Corollary~\ref{cor:one}, which says that any nontrivial $\onethird$-symmetric 3-cocycle state with $G=\ztwom$ can be reduced to $r$ disjoint copies of the Union Jack state via single-site measurements, showing that every such state is a Pauli universal resource state for MQC. We note that these single-site measurements are single-qudit, but not generally single-qubit relative to the $m$ virtual qubits at each site. However, if our aim is only to prepare a single copy of the Union Jack state, these measurements can always be chosen to be single-qubit.

Our proof is a simple index-dependent change of basis for the tensor $\tritens$, which we choose to eliminate any ``edge incidences" between some arbitrary fiducial component $\tritens(i_0,j_0,k_0)$ and all other components. More concretely, if $\tritens(i,j,k)=\tritens(i_0,j_0,k_0)=1$ are two distinct nonzero components of $\tritens$, we say these components are edge incident when two of the equalities $i=i_0$, $j=j_0$, or $k=k_0$ hold, vertex incident when only one equality holds, and disjoint when none hold. By eliminating the edge incident terms, we ensure that the measurement of all other virtual qubits in the $Z$ basis will leave a single copy of the Union Jack state on (color-specific) sites $i_0$, $j_0$, and $k_0$, up to $Z$ byproduct operators at the intersecting vertices. These byproduct operators can be accounted for with purely classical postprocessing, whereas edge-incident terms would have led to $CZ$ byproduct operators, which are nontrivial to correct for. Exhibiting such a change of basis consequently proves that our nontrivial $\onethird$-symmetric cocycle state can be reduced to one copy of the Union Jack state, while applying this procedure to each of the disjoint irreducible states in Theorem~\ref{thm:one} proves Corollary~\ref{cor:one}.

We determine our change of basis iteratively, by examining each nonzero component which is edge incident with our fiducial component $\tritens(i_0,j_0,k_0)=1$. If such a component, say $\tritens(i_0,j_0,k_1)=1$ (with $k_0\neq k_1$), is incident along the $AB$ edge, we can eliminate this component by applying a shear operation $\chi_C$ to the C index of $\tritens$. $\chi_C$ has nonzero matrix elements of $\chi_C(k,k)=1$ for all $k$ with $1\geq k\geq m$, and $\chi_C(k_0,k_1)=1$. These act on the components of $\tritens$ as $\chi_C:\tritens(i,j,k)\mapsto\tritens(i,j,k)$ when $k\neq k_1$, and $\chi_C:\tritens(i,j,k_1)\mapsto\tritens(i,j,k_1)\oplus\tritens(i,j,k_0)$ when $k=k_1$, where $\oplus$ indicates addition mod 2. It is clear that this C-site change of basis will eliminate the offending component $\tritens(i_0,j_0,k_1)$, while also avoiding the introduction of any new edge incident terms. We can utilize a similar technique to eliminate all components which are edge incident along $AC$ or $BC$ edges, showing that a series of such changes of basis will leave $\tritens$ in our desired form, with $\tritens(i_0,j_0,k_0)$ at most vertex incident with all other components. This completes our proof of Corollary~\ref{cor:one}.

\subsection{Theorem~\ref{thm:two}}
\label{sec:thmthree}

\begin{figure}[t]
  \centering
  \includegraphics[width=0.6\textwidth]{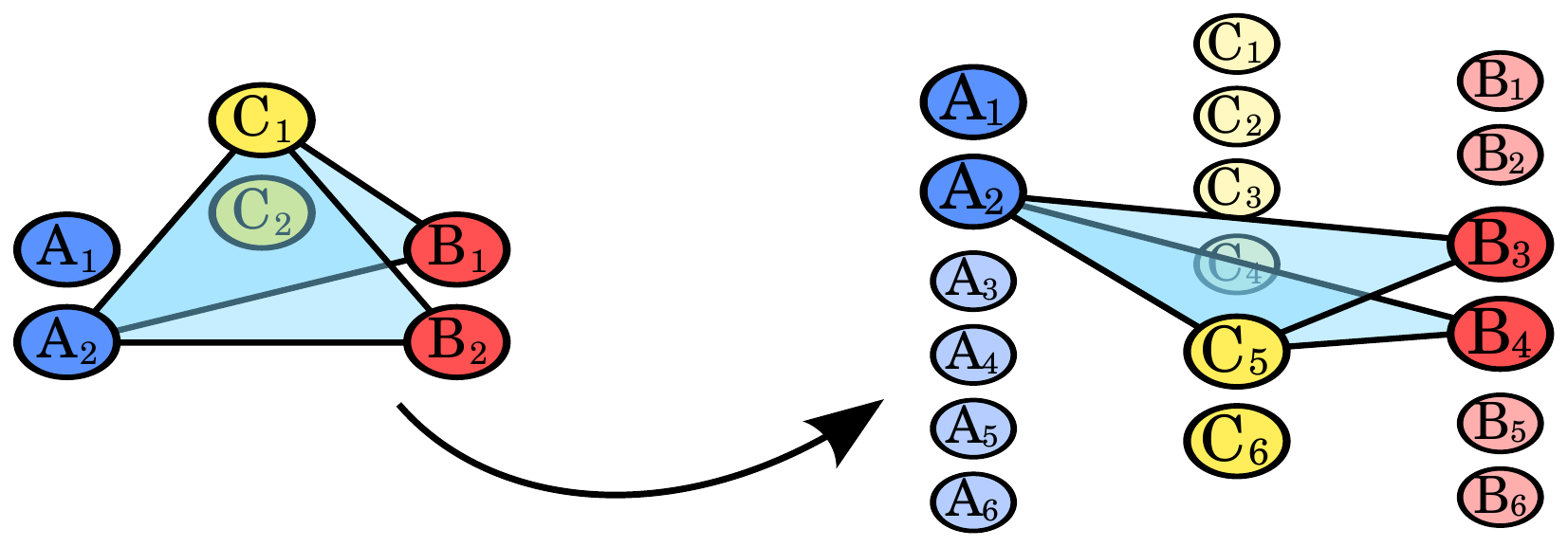}
  \caption{An illustration of our embedding technique, where our $\onethird$-symmetric cocycle state with global symmetry $G=\ztwotwo$ state is mapped to a 3-cocycle state with global symmetry $G^3=(\mathbb{Z}_2)^6$. Each lattice color is sent to a different copy of $G$ in $G^3$, so that the global application of each generator of $G^3$ acts nontrivially only at a single color of our embedded state. For example, the application of $X$ to all qubits on the fifth layer of our embedded system is equivalent to the color-dependent application of $X$ to the $C_1$ qubits of our original system. This embedding technique can be extended to achieve an embedding of constant-depth quantum circuits, which lets us show that two states in the same SPTO phase relative to fractional $\Gthird$ symmetry must be in the same SPTO phase relative to the global $G^3$ symmetry of the embedded system. The reverse implication may fail in a sufficiently general setting, owing to the lack of any obvious method for taking constant-depth quantum circuits defined on the embedded system and ``unembedding" them into circuits acting in the original setting of $|G|$-dimensional qudits.}
  \label{fig:embedding}
\end{figure}

Here we give a proof of Theorem~\ref{thm:two}, which states that every pair of non-identical $\onethird$-symmetric 3-cocycle states with $G=\ztwom$ belong to different SPTO phases relative to $\Gthird$. This in turn implies that every nontrivial $\onethird$-symmetric cocycle state has nontrivial SPTO relative to $\Gthird$. Our proof requires first describing the classification of 2D SPTO phases relative to a global symmetry $G=\ztwom$. We then use an embedding argument to show that every component of the 3-index tensor $\tritens$ itself constitutes a unique binary label of $\tristate$'s SPTO phase relative to global $G^3$ symmetry. While this doesn't necessarily give the state's SPTO phase relative to fractional symmetry $\Gthird$, an operational argument lets us show that these two classifications coincide for the case of $\onethird$-symmetric cocycle states. Consequently, the only way for two $\onethird$-symmetric states to have the same $\Gthird$ SPTO phase is to have the same component tensors, and consequently to be identical states. This suffices to prove Theorem~\ref{thm:two}.

We can use known results from \cite{propitius1995topological} (see also \cite{zaletel2014detecting}) to determine the structure of the cohomology group $\mathcal{H}^3(\ztwom,U(1))$, whose elements classify the 2D SPTO phases relative to $G=\ztwom$. For simplicity, we will write $\mathcal{H}^3(\ztwom)$ for $\mathcal{H}^3(\ztwom,U(1))$. Using a K\"{u}nneth formula, $\mathcal{H}^3(\ztwom)$ can be shown to be the direct product of groups $\mathcal{H}_{I}^3(\ztwom)$, $\mathcal{H}_{II}^3(\ztwom)$, $\mathcal{H}_{III}^3(\ztwom)$, which are respectively called type-I, type-II, and type-III factors. Each of these factors is isomorphic to a product of multiple copies of $\ztwo$, as $\mathcal{H}_{I}^3(\ztwom)\simeq\ztwom$, $\mathcal{H}_{II}^3(\ztwom)\simeq(\ztwo)^{m\choose2}$, and $\mathcal{H}_{III}^3(\ztwom)\simeq(\ztwo)^{m\choose3}$, where each $m\choose l$ indicates a binomial coefficient. By fixing a generating set for $\ztwom$, the individual $\ztwo$ components of each of these factors can be labeled by individual generators for $\mathcal{H}_{I}^3(\ztwom)$, by pairs of distinct generators for $\mathcal{H}_{II}^3(\ztwom)$, and by triples of distinct generators for $\mathcal{H}_{III}^3(\ztwom)$. Additionally, \cite{propitius1995topological} shows how one can construct model 3-cocycles for each cohomology class, which all happen to be trilinear functions.

We now show how this classification ends up giving the SPTO phases of our $\onethird$-symmetric cocycle states relative to $\Gthird$. While the $\Gthird$ symmetry of $\tristate$ isn't itself a global symmetry, we can make it into one by embedding each spin of $\tristate$, with local Hilbert space $H_G$, into a larger Hilbert space $H_{G^3}$. $H_{G^3}$ is isomorphic to three copies of $H_G$, and we choose to embed A-site spins in the first copy of $H_G$, B-site spins in the second copy, and C-site spins in the third (see Figure~\ref{fig:embedding}). The virtual qubits at each site associated with the two unused copies of $H_G$ are initialized in the $\ket{+}$ state, which ensures that the $\Gthird$ symmetry of our original state can be faithfully reproduced using the $G^3$ symmetry present in our embedded state. Recall now that SPTO phases relative to $\Gthird$ are defined operationally as equivalence classes of many-body states under the application of constant-depth, $\Gthird$-respecting local quantum circuits. By embedding these circuits into our $G^3$ setting, it is clear that any $\Gthird$-respecting quantum circuit which connects two $\Gthird$-invariant states will yield a $G^3$-respecting quantum circuit connecting the associated embedded $G^3$-invariant states. This gives us the operational result that two $\onethird$-symmetric states in the same SPTO phase relative to $\Gthird$ must be in the same SPTO phase relative to $G^3$.

Since every embedded state is itself a 3-cocycle state of $G^3$, we can use the above classification to determine the SPTO phase of any $\onethird$-symmetric $\tristate$ with respect to global $G^3$. In particular, each nontrivial component $\tritens(i,j,k)=1$ corresponds in the embedded setting to a model 3-cocycle of type-III SPTO, associated with the triple of distinct generators $(i,j+m,k+2m)$. This labeling arises from the site-dependent embedding of our original system into the larger Hilbert space $H_{G^3}$, where a generator at layer $i$ on sites A, B, or C will be sent to a generator at layer $i$, layer $i+m$, or layer $i+2m$, respectively. Because each triple is an independent label for the SPTO phase of $\tristate$ relative to $G^3$, this shows that any two $\onethird$-symmetric states with non-identical tensors $\tritens,\tritens'$ belong to different SPTO phases relative to $G^3$. The contrapositive of our operational result described above then shows that our non-identical states belong to different SPTO phases relative to $\Gthird$, which completes our proof of Theorem~\ref{thm:two}.

\end{document}